TABLE 1
Measured Linewidths of M33-like Galaxies

| | | $\Delta v$(with warp) / $\Delta v$(without warp) | | | |
|---|---|---|---|---|---|
| $i$ | $N$ | 50% of peak | 20% of peak | 50% of mean | 20% of mean |
| 0–10° | 248 | 1.14±0.05 | 1.18±0.06 | 1.31±0.09 | 1.39±0.16 |
| 10–20° | 736 | 1.08±0.09 | 1.12±0.11 | 1.16±0.14 | 1.20±0.19 |
| 20–30° | 1184 | 1.02±0.10 | 1.06±0.11 | 1.06±0.11 | 1.09±0.13 |
| 30–40° | 1592 | 1.02±0.08 | 1.04±0.08 | 1.04±0.08 | 1.05±0.09 |
| 40–50° | 1952 | 1.02±0.05 | 1.03±0.05 | 1.03±0.06 | 1.04±0.06 |
| 50–60° | 1816 | 1.01±0.04 | 1.02±0.04 | 1.02±0.04 | 1.02±0.04 |
| 60–70° | 1936 | 1.01±0.03 | 1.02±0.02 | 1.02±0.03 | 1.02±0.02 |
| 70–80° | 2280 | 1.01±0.02 | 1.01±0.01 | 1.01±0.02 | 1.01±0.01 |
| 80–90° | 2064 | 1.00±0.01 | 1.00±0.01 | 1.01±0.01 | 1.01±0.01 |
| All | 13808 | 1.02±0.06 | 1.03±0.07 | 1.04±0.08 | 1.05±0.10 |

# A WARPED DISK MODEL FOR M33 AND THE 21-cm LINE WIDTH IN SPIRAL GALAXIES

Edvige Corbelli[1] and Stephen E. Schneider[2]

Preprint n. 26/96

[1]Osservatorio Astrofisico di Arcetri,
 Largo E. Fermi 5, I-50125 Firenze (Italy)

[2]Five College Astronomy Department and Department of Physics and Astronomy,
 University of Massachusetts 632 Lederle Tower, Amherst, MA 01003

**ABSTRACT**

To determine the actual HI distribution and the velocity field in the outermost disk of the spiral galaxy M33, a tilted-ring model is fitted to 21-cm line data taken with the Arecibo Telescope. Since M33 is one of the main calibrators for the extragalactic distance scale derived through the Tully-Fisher relation, the outer disk warping is of interest for a correct determination and deprojection of the galaxy's line width. Even though our best model predicts small effects on the observed line width of M33, we show that similar outer disk warping in galaxies oriented differently along our line of sight could affect the widths considerably. Therefore there may be systematic effects in the determination of the rotation velocities and dynamic masses of spiral galaxies, whose exact value depends also on which method is used for measuring the galaxy's total line width.

Subject headings: ISM: HI — galaxies: warps, extragalactic distance scale

# 1. Introduction

M33 (NGC 598) is the largest spiral galaxy in angular extent that can be observed with the Arecibo 305-m radiotelescope, (The Arecibo Observatory is part of the National Astronomy and Ionosphere Center, which is operated by Cornell University under cooperative agreement with the National Science Foundation.) making it ideal for studying the kinematic and neutral hydrogen structure of an outer disk down to a very faint level. A previous study of the HI emission in M33 (Corbelli, Schneider, & Salpeter 1989, hereafter Paper I) showed that the HI disk starts tilting outside the optical image of the galaxy. However, the partial coverage of the HI extent in Paper I (using a beam of 3.9 arcmin we observed points along a hexagonal grid with 9 arcmin spacings) did not allow us to determine the effective inclination of the outer disk at each radius or to draw firm conclusions about the shape of the rotation curve and the extent of the galaxy's massive halo in the outermost parts. The inclination changes also affect interpretations of the sharp HI fall-off in total HI column density that we observed in Paper I, and this is important indirectly for studying the UV and soft X-ray background (Corbelli & Salpeter 1993; Maloney 1993; Dove & Shull 1994).

We have subsequently carried out more thorough and higher sensitivity observations of the HI in M33, which are the subject of this paper. A more comprehensive understanding of the HI in M33 is also motivated by M33's use as a primary calibrator of the Tully-Fisher relation (Freedman, Wilson, & Madore 1991; Pierce 1988). The Tully-Fisher relation between the 21-cm line width and the absolute magnitude of a galaxy (Tully & Fisher 1977) has become one of the most widely used extragalactic distance indicators, so it is important to understand how the warping that we observe in M33 might affect distance estimates. There are several issues related to warping in M33 and other galaxies which we will try to address in this paper: (1) Is the line width of M33 being properly corrected for inclination? (2) If other galaxies have similar warps in their outer disks, what would be the probable errors in the attempting to determine their true rotation speeds from their observed line widths? (3) What is the optimum method and intensity level for measuring the line width which minimizes the effects of the warping?

In Section 2 we describe the new observations, which include a search for HI islands in the area surrounding M33. In Section 3 we describe the procedure for modeling the warping in the outer disk. The main results of the fit are outlined in Section 4 where we also discuss the possible influence of M31. In Section 5 we investigate how a warped HI disk like that in M33 affects distances determined by the Tully-Fisher relation and in Section 6 we see if such effects are visible in the line widths of available HI data. Finally, in Section 7 we summarize our results.

## 2. The HI Observations

The HI distribution in the optical region of M33 has been previously studied in detail by Deul and van der Hulst (1987), and the surrounding areas have been mapped by Huchtmeier (1978) and in Paper I. We have carried out further observations with the Arecibo 305-m radiotelescope using the same system and procedures as in Paper I, but with better coverage of the entire HI extent.

A total of 611 spectra were taken using the 21-cm linear polarization "flat" feed over an hexagonal grid with 4.5 arcmin spacing. During the observing run, pointing errors were found to be smaller than 30 arcsec for continuum sources at zenith angles like those for M33. The limiting sensitivity for HI detected in these observations was typically 1–2 Jy km s$^{-1}$beam$^{-1}$, corresponding to an HI column density of 1–2×10$^{19}$ cm$^{-2}$.

Further information about the antenna reception pattern, sensitivity and sidelobes are given in Paper I, as well as technical details of the observations and data reduction of the individual spectra. We also discuss problems of beam asymmetry in Paper I; this is particularly important for measuring regions where the column density is rapidly changing as were studied in that paper. The beam asymmetries are of minor significance for the modeling discussed in this paper because variations in the radial beam efficiency amount to at most a few percent, while our models are attempting to fit at a much less stringent level.

The sum of all the observations yields an effective sensitivity that is nearly uniform over the entire galaxy. The total flux found this way ($2.1 \times 10^4$ Jy km s$^{-1}$) must be corrected by the ratio of the area per beam on the grid (17.5arcmin$^2$) to the effective beam area ($\sim 23$ arcmin$^2$) integrated out to 30' for these relatively large zenith angle observations. This gives a total integrated flux of $1.6 \times 10^4$ Jy km s$^{-1}$ for M33, or an integrated HI mass of $1.8 \times 10^9 M_\odot$ assuming M33 has a distance of 0.69 Mpc. Our flux is at the upper end of previously reported measurements; fluxes reported in papers listed in the bibliographic catalog of Huchtmeier & Richter (1989) range from 0.7 to 1.6×10$^4$ Jy km s$^{-1}$. We believe our value represents the best global determination to date based on the extent and sensitivity of our mapping, but given the potential systematic uncertainties due to factors like the calibration, beam area, zenith angle and frequency response corrections we estimate an uncertainty of up to ∼20%.

The integrated HI profile of M33 (Fig. 1), like that of most galaxies, shows some asymmetry, with about 5% more flux in the high-velocity half of the profile. This is quite symmetric compared to typical galaxy HI profiles (Richter & Sancisi 1994). At 50% of the peak flux densities in each "horn" the mean velocity is −181 km s$^{-1}$; and the same value is measured at 20% and 80% of the peaks. However, because of the excess of gas at higher velocities, the flux-weighted mean velocity is −179 km s$^{-1}$; this difference is small but proves to be significant in our fitting procedure discussed later. At 50% of peak the line width is 190 km s$^{-1}$, while at 20% and 80% of the peak, the widths are 214 and 174 km s$^{-1}$ respectively. The 20% value may be affected by slight confusion with the Milky Way on the high-velocity edge, although we have attempted to carefully excise this in the individual spectra. Again, values reported in the papers listed in the Huchtmeier & Richter catalog vary from these measurements, with widths that are generally smaller.

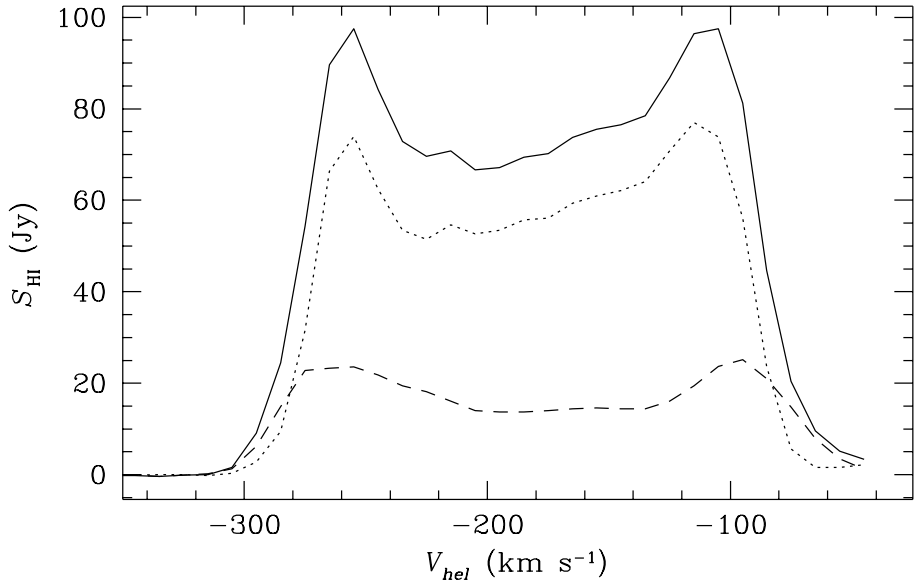

**Figure 1.** The 21 cm HI profile of M33 (solid line). The dotted line shows the contribution from the area within the 25 mag arcsec$^{-2}$ isophote, and the dashed line from the regions exterior to this.

The range of fluxes and line widths reported for M33 in the literature appear to be caused by the differences in the inner and outer portions of the galaxy in combination with the beam sizes of the different telescopes used to map it. We show in Fig. 1 the contributions from inside and outside the bright disk as defined by the $71 \times 42$ elliptical boundary of the 25 mag arcsec$^{-1}$ isophote. 74% of the HI emission comes from the bright disk, and has a mean velocity of $-179$ km s$^{-1}$ and a line width at 50% of peak of 187 km s$^{-1}$(207 km s$^{-1}$at 20%). The outer regions have a mean velocity of $-180$ km s$^{-1}$ and a larger line width: 217 km s$^{-1}$(240 km s$^{-1}$at 20%). Limited mapping, and measurements made with interferometers would probably miss much of this extended HI, which could explain the smaller fluxes and narrower line profiles.

Results of the detailed survey of the outer disk indicate that the HI there is not very regular in its extent and density structure. As we shall discuss later, it is possible that this is partially due to the proximity of M31. However, it should be noted that even isolated galaxies can show quite irregular HI distributions in their outermost parts (see for example NGC 3344 in Paper I or NGC 3198 in Begeman 1989). These irregularities limit our ability to search for effects like non-linear orbits, radial motion, etc. If the residuals of a simple circular rotation model do not follow a symmetric pattern, they more likely result from underlying local irregularities or perhaps external disturbances like ram pressure stripping.

In the course of these observations we also carried out some high sensitivity observations to determine if there might be very low column densities of HI outside of the regions mapped in Paper I. We briefly summarize the negative results of these searches here: We used the dual circular polarization feed, which is much more sensitive than the flat feed, but which suffers from much higher sidelobes. We searched for HI north of M33 in an area covering 60 arcmin in right ascension and 30 arcmin in declination (around $1^h$ $27^m$ $-32°$ ), just outside the region where the sharp HI fall-off in the outer disk was mapped in Paper I. No HI signals were detected in spectra having a resolution of 4.1 km s$^{-1}$ with a single-channel uncertainty of $3\sigma \sim 6$ mJy. This implies there is no HI at a column density of $\sim 3 \times 10^{18}$ cm$^{-2}$ for a supposed signal width of 50 km s$^{-1}$. In similar searches to the northwest and southeast of M33 (around $1^h\ 26^m$ –31° and $1^h\ 35^m$ –30°) we cannot exclude a faint level of HI as high as $6 - 8 \times 10^{18}$ cm$^{-2}$. But since these observations were carried out with the circular feed and the telescope is no longer pointing along the major axis, this flux might be caused by distant sidelobe contamination from HI already located closer in to M33.

## 3. Modeling Procedures

To represent the overall distribution of HI, we use a large number of tilted concentric rings in circular rotation around the center as in Bosma (1981). Each ring is characterized by a fixed value of the HI surface density $\Sigma_{HI}$, circular velocity $V$, inclination angle $\phi$ and position angle of the major axis $\theta$. To carry out tests on the model, each ring is subdivided into segments of equal area that are small compared to the telescope beam area. For each segment we compute the peak velocity along the line of sight given $V$, $\phi$, and $\theta$ for that ring and we assign a velocity dispersion. We assume the gas in each ring segment is characterized by a Gaussian emission line of width $w$ around $V(r)$, whose integrated emission equals $\Sigma_{HI}(r)$. $w$ is assumed constant and isotropic over the whole galaxy. Initially we set $w = 12$ km s$^{-1}$, but $w$ as well as the systemic velocity of the galaxy are thereafter considered as free parameters.

It has been shown that the vertical component of the velocity dispersion in some nearly face-on galaxies does not vary much between the inner and outer disks when star formation is absent (see for example Dickey, Murray & Helou 1990). However, the presence of high velocity gas and turbulence due to star formation in the inner disk can give a higher velocity dispersion than in the outer disk (Kamphuis 1993). In addition, the radial and tangential components of the velocity dispersion are expected to be about 1.5 times larger than the z-component, but we shall show later that M33's inclination is relatively constant everywhere, so the relative contribution of the different components remains the same (see Bottinelli et al. 1983). The velocity dispersion may also vary with radius, and there may be deviations from circular motion. These complications are extremely difficult to disentangle uniquely from more basic model parameters or from observational variations, so for the present paper we take the basic approach of considering only a single-value of the velocity dispersion everywhere.

Simulating an individual observation also requires a model of the directional sensitivity of the telescope beam. By determining the relative contribution of various points of the galaxy within the main beam and first sidelobes, our model will simulate effects like "beam smearing." The beam shape can be well matched by a beam function $f_b$ which is the sum of three Gaussian functions:

$$f_b(r) = e^{-\frac{r^2}{2\sigma_1^2}} + \frac{0.15\sigma_1^2}{0.85(\sigma_2^2 - \sigma_3^2)}\left[e^{-\frac{r^2}{2\sigma_2^2}} - e^{-\frac{r^2}{2\sigma_3^2}}\right] \quad (1)$$

with $\sigma_1 = 1'.8$, $\sigma_2 = 4'.5$, and $\sigma_3 = 3'.5$. The integrated beam area of this model gives a good match to the radial beam efficiency of the Arecibo flat feed as described in the Appendix of Paper I with 80–85% of the integrated beam area in the main beam. However, we have enlarged the main beam to 4'.15 (FWHM) which is the mean size at the average $\sim 15°$ zenith angle of these observations.

Finally, we calculate model fluxes at each position where HI was observed in M33, and test how well the spectra match. To produce the spectra we calculate the spatial sum of all the individual ring segments multiplied by the beam function in their direction. We determine the model flux densities in 28 velocity channels that span the velocity range over which emission is observed. Each channel is 10 km s$^{-1}$ wide, and they cover –40 to –320 km s$^{-1}$ (heliocentric). At each map position we compare the spectrum derived from the model with the measured spectrum. In our models we set the number of rings to 110, the width of each ring to 0.7 arcmin, and the surface area of the ring segments to 1.0 arcmin$^2$.

We start with a ring model in which the inner disk has an HI inclination and position angle as derived by the fit of Rogstad, Wright & Lockhart (1976). The outer disk properties are estimated from our data, and roughly agree with Rogstad et al., although their data extend only about half as far out and their inclination at their outermost radius is much steeper than our data indicate. This starting model can be improved by examining the symmetry properties of the velocity field. As shown by Warner *et al.* (1973) and by van der Kruit & Allen (1978), errors in the parameters produce characteristic patterns in the residual velocity field. Therefore at each point of the grid where HI was detected we calculate the difference between the observed mean velocity, $V^{obs}$, and the mean velocity predicted by the model $V^{mod}$. We also compute the ratio between the integrated flux observed at each position, $I^{obs}$, and the integrated flux predicted by the model $I^{mod}$. The intensity and signs of the deviations from the predicted values were plotted for both the velocities and intensities, and we interactively adjusted the parameters to find a "reasonable fit" with small residuals and avoiding systematic symmetries around the rings. The "reasonable fit" becomes the starting point for a minimum $\chi^2$ procedure, where we vary the model parameters to find the best fit.

The assignment of a measure of goodness of fit is the critical step in this sort of modeling procedure, and we depart here slightly from previous schemes. Rather than fitting to the moments of the flux distribution, we compare our results directly against the full spectral database. This consists of the fluxes $I_{n,i}$ at each of the $n$ (611 total) positions and in each of the $i$ (28 total) channels. We minimize the differences between the observed and modeled fluxes in each spectral channel at all of the observed positions of the grid in order to derive a "shape error." This procedure avoids the difficulty of assigning uncertainties to quite disparate quantities like fluxes, velocities, and velocity dispersions. It also retains information about the line shape that are lost when just the first few moments are examined, like, for example, in the regions of M33 where there is a bimodal velocity distribution of the gas.

For spectrum number $n$ we estimate that the experimental uncertainty for each channel is:

$$\sigma_n = 2 \frac{\Delta v \times rms_n}{\sqrt{\Delta v / 4.1 \text{ km s}^{-1}}} \ , \tag{2}$$

where $\Delta v$ is the width of the channel (interpolated to 10 km s$^{-1}$ for these data), and $rms_n$ is the measured standard deviation about the baseline in the original spectra (which had a resolution of 4.1 km s$^{-1}$ after Hanning smoothing). We multiply the uncertainty by 2 to take into account possible baseline and pointing errors.

We initially tried a solution based on this channel-by-channel scheme alone, however it significantly underestimated the total flux. We believe this is because the uncertainties are large compared to the channels with weaker fluxes so that even zero fluxes can be within the model tolerances. Therefore we include an additional "flux error" term in our $\chi^2$ calculation, which compares the integrated fluxes ($\sum_i I_{n,i} = I_n$) of each observed spectrum to that generated by the model. The flux term is also affected by the experimental uncertainties given in equation (2) above (with $\Delta v = 280$km s$^{-1}$), but it is primarily influenced by the calibration uncertainties, which are proportional to the flux. This flux error term is also useful in forcing the solutions to "pay more attention" to the warp and outskirts of M33, since an error proportional to the flux can force the minimization to be almost as sensitive to differences between the model and observations in weak-line regions as in the inner high-flux regions. We estimate that the calibration errors are probably of order 10–15% of the flux, but to tighten the constraints on the model we adopt a smaller uncertainty of 5%.

In our final scheme, we derive the shape error by normalizing each model spectrum $n$ to have the same total flux as the data, and we then find the sum of the individual channel errors relative to this normalized model. This yields an error in the modeled spectral shape that is relatively independent of calibration errors or local variations in the HI surface density. Our resultant reduced $\chi^2$ formula is formed by adding the flux and shape terms:

$$\bar{\chi}^2 = \frac{1}{N - N_p} \sum_{n=1}^{N} \left[ \frac{\left(I_n^{mod} - I_n^{obs}\right)^2}{28(\sigma_{n,i})^2 + (0.05 I_n^{obs})^2} + \frac{1}{28} \sum_{i=1}^{28} \frac{\left(\frac{I_n^{obs}}{I_n^{mod}} I_{n,i}^{mod} - I_{n,i}^{obs}\right)^2}{(\sigma_{n,i})^2} \right] \quad (3)$$

where $N$ is the number of positions (611), and $N_p$ is the number of parameters (16 for our basic model), leaving $\sim 600$ degrees of freedom. If we correctly estimated the uncertainties in our measurements and our model were perfect, the reduced $\chi^2$ measurement for the flux and shape terms would each have a minimum value of order unity for this large number of degrees of freedom, so our minimization would ideally reach $\bar{\chi}^2 \approx 2$. However, since we have chosen a low calibration error (5%, to better constrain the solution), we expect a best reduced $\chi^2$ value higher than unity. And, of course, since a ring model cannot generate asymmetric distributions or local variations, we expect the minimum $\bar{\chi}^2$ value to be even higher. Our procedure is to search for a minimum and to compare how well different models do. As a reference point, we note that the unmodeled fluxes (setting the intensities in the model to zero) yield a raw value of $\bar{\chi}_0^2 = 230$.

Given the difficulty in finding a unique minimum $\bar{\chi}^2$ solution when there are so many free parameters, we use two methods to converge towards the minimal solution. In the first method, we set starting values for each parameter according to the "reasonable fit" described above, giving a permissible range of variation for each parameter. Since some of the parameters might be correlated we begin by searching for minima over a grid of the parameters surrounding the reasonable fit. We evaluate $\bar{\chi}^2$ for all the possible combinations of parameter values, their starting, maximum, and minimum values. After iterating to smaller ranges of variation, in the end we choose the parameter values which give the minimum $\bar{\chi}^2$. In the second method we apply a technique of partial minima. We evaluate the $\bar{\chi}^2$ by varying each parameter separately and interpolating to estimate the value for a minimum $\bar{\chi}^2$ for each parameter, and then repeat the procedure over smaller parameter intervals around the new solution.

## 4. Models of the HI Distribution

### 4.1) The Basic Model

There are a very large number of degrees of freedom if we use a model in which every ring varies freely, and the solutions are not very robust (see Section 4.3). However, Briggs (1990) showed that for a dozen galaxies with deep HI measurements (including the M33 data of Rogstad et al. 1976), the warps are generally characterized by a gradual shift at around the Holmberg radius to an outer, relatively-fixed orientation. These data for M33 did not extend as deep as ours and did not show a fixed outer orientation, but based on an initial examination of our data it appeared plausible to postulate a model of this form. We describe this "basic model" in terms of hyperbolic tangent functions of the radius $r$ (in arcmin) of the following forms:

$$
\begin{aligned}
V(r) &= V_\infty \tanh(r/\Delta_V) & \text{km s}^{-1} \\
\Sigma_{HI}(r) &= \Sigma_0 + \frac{1}{2}(\Sigma_\infty - \Sigma_0)\{1 + \tanh[(r - R_\Sigma)/\Delta_\Sigma]\} & \text{Jy km s}^{-1} \\
\phi(r) &= \phi_0 + \frac{1}{2}(\phi_\infty - \phi_0)\{1 + \tanh[(r - R_\phi)/\Delta_\phi]\} & \text{deg} \\
\theta(r) &= \theta_0 + \frac{1}{2}(\theta_\infty - \theta_0)\{1 + \tanh[(r - R_\theta)/\Delta_\theta]\} & \text{deg}
\end{aligned}
\qquad (4)
$$

The parameters subscripted 0 and $\infty$ represent the central and limiting outer values of the parameters while the $R$ and $\Delta$ parameters indicate the radius and range over which the particular variable changes. The form of the velocity in equation (4) gives a flat outer rotation curve and a linear rise at the center over the range $\sim \Delta_V/2$. The other functions step from a relatively constant inner value to a fixed outer value, although by choosing large values of $R$ and $\Delta$, the tanh functions allow for solutions where the variables change continuously in the outer regions of the galaxy. We will discuss later the results for a variety of refinements of the model.

Our functional form for $\Sigma_{HI}$ is a little unusual in that it does not include an exponential decline in the HI surface density as is often assumed. In fact, it is clear from the observations that the HI has a local minimum at the center of the galaxy and is relatively constant over the bright disk of M33 (see for example Deul & van der Hulst 1987). In the outer disk, the surface density declines rapidly (see Paper I), but there are low levels of gas detected to quite large radii in some regions; the radial distribution found by Rogstad et al. (1976), for example, might be described by a tanh function. We use the tanh function primarily to model the inner gas and decline, and we use the small outer surface density of gas represented by $\Sigma_\infty$ to match locations where gas is actually detected, although the real outer distribution is clearly clumpier than can be represented by a simple ring model. Because the model is so simple, it is primarily useful for describing the basic differences between the inner and outer disks.

After carrying out the $\bar{\chi}^2$ procedure described in the previous section, the parameters of the basic model are found to be:

$$
\begin{aligned}
V_\infty &= 107.1 \pm 2.9 \quad \Delta_V = 8.47 \pm 1.75 \\
\Sigma_0 &= 95.9 \pm 11.7 \quad \Sigma_\infty = 3.41 \pm 1.31 \quad \Delta_\Sigma = 11.9 \pm 2.0 \quad R_\Sigma = 29.6 \pm 1.5 \\
\phi_0 &= 49.0 \pm 3.5 \quad \phi_\infty = 56.8 \pm 2.2 \quad \Delta_\phi = 3.37 \pm 3.37 \quad R_\phi = 23.4 \pm 3.8 \\
\theta_0 &= 21.1 \pm 2.0 \quad \theta_\infty = -13.0 \pm 7.8 \quad \Delta_\theta = 7.00 \pm 3.75 \quad R_\theta = 40.8 \pm 2.0
\end{aligned}
\tag{5}
$$

which give a reduced $\chi^2$ value of $\bar{\chi}^2 = 23.7$. The functions $V$, $\Sigma$, $\phi$, and $\theta$ are plotted in Fig. 2. In the basic model, the surface density of HI drops by a factor of 30 from the inner to the outer regions and the velocity rises to an almost constant value in about 20 arcmin, well within the optical radius of M33. The rotation speed of the disk, 107 km s$^{-1}$, matches the value found by Reakes & Newton (1978).

The error intervals we list correspond to the range over which the value of $\bar{\chi}^2$ increases by an amount that would imply a $1\sigma$ (68.3%) probability interval. For a large number of degrees of freedom, $N$, $\bar{\chi}^2$ has a standard deviation of $\sqrt{2/N}$. For the flux term in equation 3 this corresponds to a variation in $\bar{\chi}^2$ of $\sim 0.058$, and for the shape term $\sim 0.011$. The basic model is primarily constrained by the flux term, but even so, a variation of only 0.058 in $\bar{\chi}^2$ would imply improbably small errors. Since the constraints on the basic model do not allow it to approach the theoretical minimum $\bar{\chi}^2$, we treat it as a relative measure and solve for the range over which each variable results in a 5.8% increase $\bar{\chi}^2$. Since each variable is tested independently, the errors are only indicative since the effects may be correlated.

All of the parameters, except $R_\phi$, are well-behaved in the sense that the $\bar{\chi}^2$ is nearly parabolic around our minimum solution. $R_\phi$ is not very well constrained because the inclination of the galaxy, $\phi$, is almost the same from the inner to the outer disk, so that the warping axis is nearly along the line of sight. Therefore the radius and range over which $\phi$ changes are fairly weakly constrained; a good fit is obtained even setting $\phi$ equal to a constant. This contrasts strongly with the model of Rogstad et al. (1976), which has the outer disk becoming almost edge-on at a radius of only $\sim 45$ arcmin.

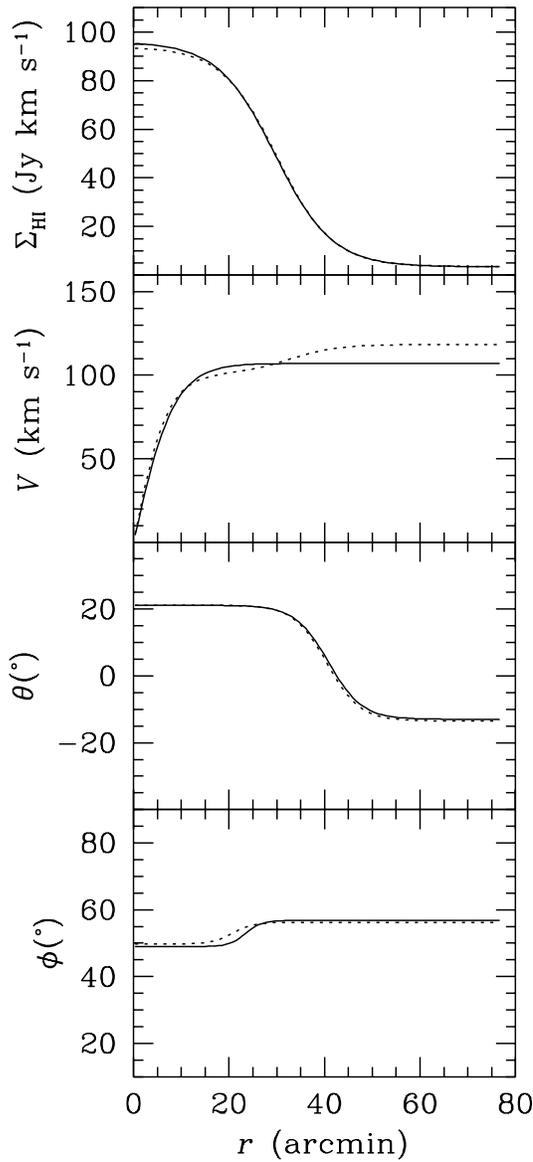

**Figure 2.** The basic model of M33. The solid lines show the best-fit results described by equation (5), while the dotted line shows results for a model with a second step in the velocity described by equation (7).

The actual angle between the inner and outer disks can be calculated from spherical trigonometry:

$$\cos\psi = \cos\phi_\infty \cos\phi_0 + \sin\phi_\infty \sin\phi_0 \cos(\theta_\infty - \theta_0)$$

which in this model corresponds to 28°.

We also included the systemic velocity and the velocity dispersion in the minimization, and find $V_{sys} = -179.2 \pm 1.7$ and $w = 12.9 \pm 1.8$. We use an isotropic dispersion in our models, but if we were to follow the procedure of Bottinelli et al. (1983) and assume that the radial and tangential components are each 1.5× the $z$-component, the observed velocity dispersion would be related to z dispersion by:

$$w_{obs} = \left[(w_z \cos \phi)^2 + (1.5 w_z \sin \phi)^2\right]^{1/2} . \tag{6}$$

Since the inner and outer disks are both at nearly the same inclination, the ratio $w_{obs}/w_z$ would be nearly constant everywhere, and for our case is $\simeq 1.24$. Under these assumptions, our best-fit dispersion would imply a z-dispersion of approximately 10.4 km s$^{-1}$. This value is appropriate for comparison to other studies of the face-on dispersion, but since the data are not sufficiently detailed to permit us to directly determine the value, we continue to use an isotropic dispersion in our further modeling of M33.

In Paper I we detected a sharp HI fall-off in the outermost disk of M33. This starts at about 2 optical radii, where the HI column density along the line of sight drops below $\sim 3 \times 10^{19}$ cm$^{-2}$ (beyond this fall-off no HI has been detected to a level of $10^{18}$ cm$^{-2}$). According to the basic model described in this Section, the HI column density perpendicular to the galactic plane where the sharp HI fall-off occurs is therefore of order $\sim 2 \times 10^{19}$ cm$^{-2}$.

*4.2) Comparison of the Model and Data*

The integrated HI profile from the basic model is plotted in Fig. 3 and compared to the actual profile. We also show the contributions from inside and outside the bright inner region corresponding to the same regions as shown in Fig. 1. The minimum $\chi^2$ solution is skewed to a slightly more positive velocity (mean velocity of $-179$ km s$^{-1}$) than the value calculated from the profile edges, but it closely matches the flux-weighted velocity of the entire galaxy. This might be expected for a model based on flux comparisons.

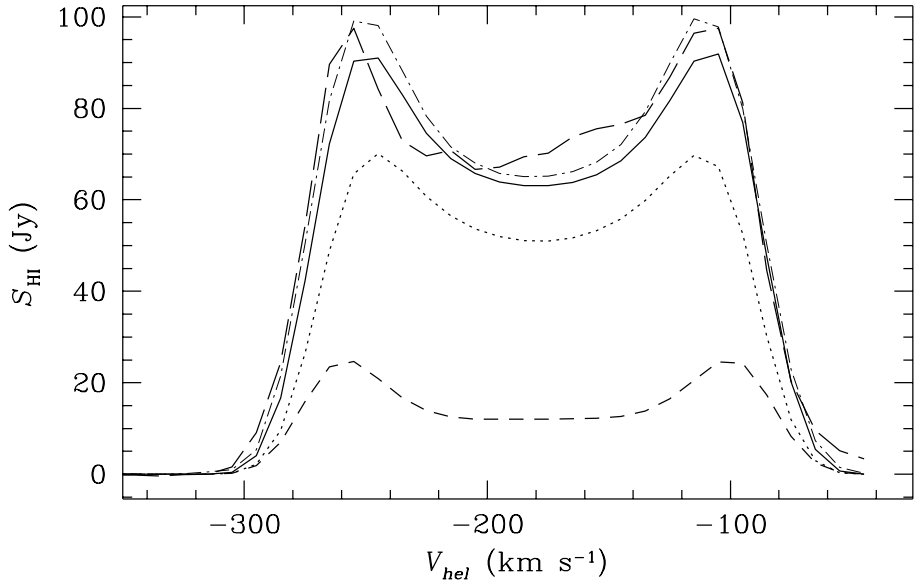

**Figure 3.** Comparison of the HI profile derived from the basic model (solid line) with the observed profile (long dashes). The contributions from inside and outside the 25 mag arcsec$^{-2}$ isophote are shown as in Figure 1. The dot–dash line shows the total profile for the "free-ring" model described in Section 4.3.

Forcing the model to have the same mean velocity as the edge measurements of the observed profile raises the value of $\bar{\chi}^2$ to 26.2 and only slightly changes the other parameters. The line widths both for the total profile and for the inner and outer disks all agree with the observations to within 5km s$^{-1}$.

The minimum $\chi^2$ solution underestimates the total HI mass of M33 by about 6%. This may be caused by several factors. First, because we are minimizing $\bar{\chi}^2$ in a simple linear sense when there are multiplicative errors, the $\chi^2$ routines tend to be biased in favor of smaller fluxes; for example, twice the flux will yield a larger $\chi^2$ than half the flux even though both would correspond to the same size calibration error. For a similar reason, lower-flux regions modeled by a ring exert more influence on the solution, so asymmetries within the HI distribution tend to cause lower fluxes. Finally, a small portion of the excess in the measured flux may also be due to Galactic contamination that appears as a weak unmodeled excess in the high velocity channels (between $-80$ and $-40$ km s$^{-1}$).

A detailed comparison of the differences between the basic model and the data is shown in Fig. 4. Here we show how the integrated fluxes and flux-weighted mean velocities differ at each point in the observed spectra and the synthesized spectra. Plus and minus symbols are shown with sizes proportional to the differences (except for a relatively small number of outliers which are plotted at the maximum symbol size). The projections of the model rings on the sky at ten radii are also shown.

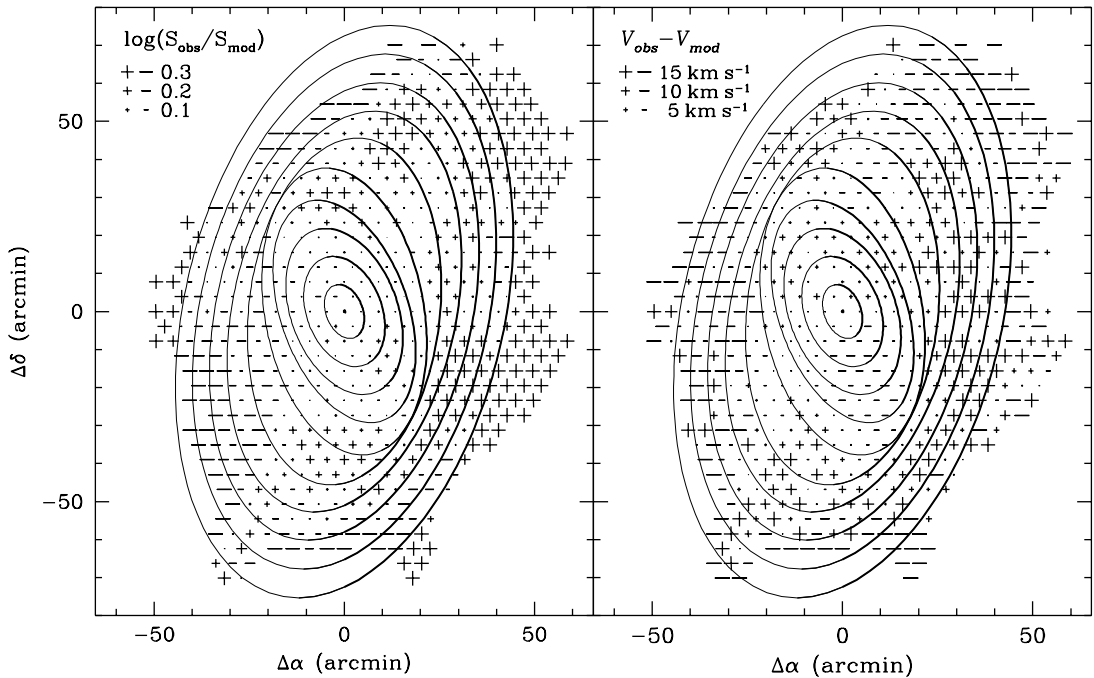

**Figure 4.** Maps of flux and velocity residuals for the basic model. (a) The residuals of the integrated flux per beam are shown at the position of each measurement with a plus or minus sign whose size is proportional to the log-ratio of the observed and modeled results. Symbol sizes corresponding to three levels of residuals sizes are shown for reference. (b) The residuals of the flux-weighted velocity at each point are shown with a plus or minus sign sized in proportion to the velocity difference

Overall, the residuals are generally small within the bright inner disk, and much larger and more variable in the outer disk. The flux residuals show that the outer disk is asymmetric. The HI extends beyond the model to the northwest, while the fluxes to the southeast are weaker than in the model. The velocity residual pattern could indicate some radial streaming motions, but, as we find with the more complex models studied below, a similar pattern can be generated by changes in the position angle of the orbiting material.

There is also a large-scale pattern to the velocity residuals in the inner disk. The mean velocities are generally more positive than in the model on the north (approaching) side of the galaxy and more negative in the south. This is mainly an artifact of showing flux-weighted mean velocities in the figure; these tend to be biased by the integrated flux in far sidelobes and noise fluctuations toward a velocity closer to the mean of the galaxy. This problem can become even worse in the outer disk where the signals are sometimes so weak that the mean velocity is very poorly determined. This problem does not affect our minimization procedure as much because the flux differences in channels well-removed from the predicted velocities have little "leverage" on the solutions.

On a finer scale, there are some peculiar patterns in the residuals that suggest additional complexities in the HI distribution. Northeast of the inner disk there is a fairly narrow strip of high flux points running almost perpendicular to the bright disk and extending across the entire mapped region. This appears to have a corresponding peculiarity in the velocity field, and some of the individual spectra through this region have two peaks. There is a less well defined excess of fluxes symmetrically placed on the south side of the galaxy with velocity peculiarities of the opposite sign. This pattern is suggestive of a distinct outer ring, which could be an extension of the loop of HI seen in the northern extreme of the outer disk (Paper I). We will examine this possibility further below.

Our basic model is not ideal, but it does quite well considering its simplicity. It is a good description of the basic differences between the inner and outer disks, but more complex models or a better understanding of a possible interaction with M31 could presumably decrease the residuals. We examine some of these possibilities next.

### 4.3) More-Complex Models

As a first attempt to model the HI distribution in more detail, we modified the functional forms of $\Sigma$, $V$, $\phi$, $\theta$ in the outer disk, allowing an additional tanh function to provide another step in the values at an additional radius. No significant improvement in the $\chi^2$ value was found when adding shifts to $\phi$, $\theta$, and $\Sigma_{HI}$. However, a small improvement was found when a shift in $V$ was added, yielding the following velocity function:

$$V(r) = 101.2\tanh(r/7.2) + \frac{1}{2}(118.2 - 101.2)\left[1 + \tanh\left(\frac{r - 32.6}{8.7}\right)\right] \text{ km s}^{-1} . \quad (7)$$

This would indicate a rotation speed that is slightly smaller in the bright disk than the basic model, but which rises to 118 km s$^{-1}$ in the outer disk of M33. The minimum $\bar{\chi}^2$ solution for this model is plotted in Fig. 2 with a dotted line. The value of $\bar{\chi}^2$ for this model is 22.5, which is not a dramatic improvement, but it argues against the presence of a decrease in the rotation speed at large radii. Thus, the total mass appears to grow at least $\propto R$ out to more than twice the optical radius.

It is obvious that a symmetric model is insufficient to address the asymmetries and peculiarities in the HI distribution described in Section 4.2. Therefore, we next attempted to model M33 with rings that follow no preset functional form. We constructed a model in which the properties at eleven equally spaced radii could be varied independently; the properties for ten rings between each of these radii were then linearly interpolated. All of the ring properties were allowed to vary, including even the positions and systemic velocities of the ring centers.

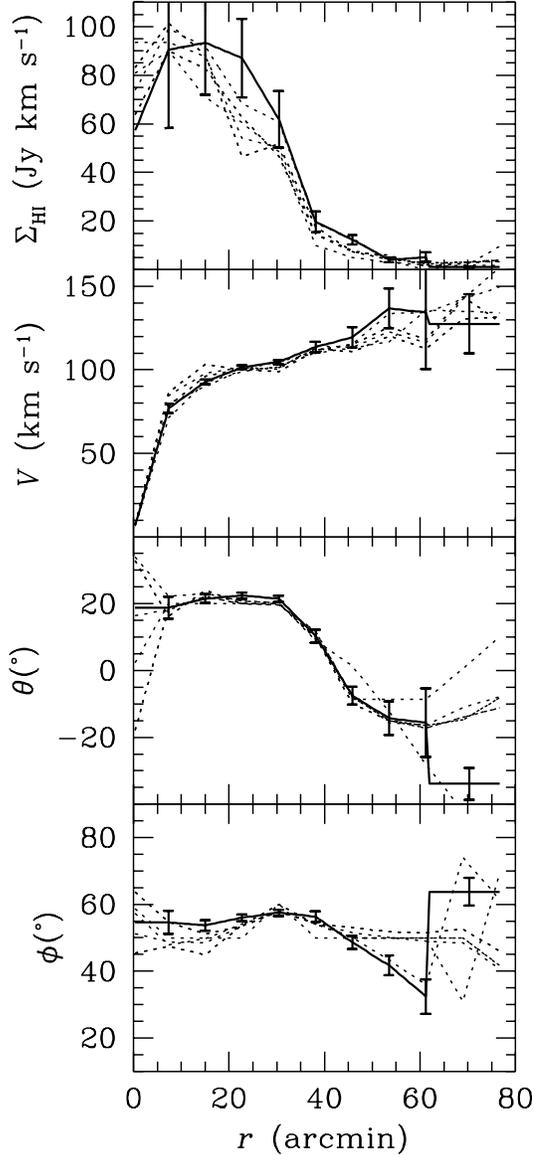

**Figure 5.** An assortment of "free ring" models fit to the HI data, as in Fig. 2. The minimum $\bar{\chi}^2$ solutions depend on initial parameter choices and the method of finding the minimum. The dotted lines indicate the range of results found. The solid line with indicative error bars represents the result of the preferred model developed in the text.

This type of model is difficult to manage because of the large number of parameters (77) and the high degree of degeneracy resulting from variations of many of them. We carried out several optimization attempts under a variety of initial conditions and with different orderings for adjusting the parameters. We found that the results for the were not always robust. With the freedom to shift rings in position and velocity, the minimization procedure sometimes locked into improvements of a single region of locally large $\chi^2$ values, yielding quite different properties. We display an assortment of low-$\bar{\chi}^2$ ($10 < \bar{\chi}^2 < 15$) functional solutions that were found in Fig. 5.

Even though these fits still have a very large number of degrees of freedom and smaller $\bar{\chi}^2$ values, we cannot assert with any confidence that they represent well-defined solutions. One of the main problems in these models was their instability toward forming highly deviant outer rings because of the *lack* of observations in very distant regions. The portions of a ring that fall entirely outside the observed region could wander about with little constraint. Also, at the center of the galaxy the line width of the HI profile is larger than any of the models predict—most likely because of a larger velocity dispersion in the galaxy center—and the fits are unconstrained. The intermediate regions, where the solutions in Fig. 5 are similar, appear to be more robust, as a more formal analysis indicates below.

To stabilize the free-ring model against "wandering outer rings," we surrounded the galaxy with artificial zero-flux observations. We also find it more stable to first determine a solution for the central positions and velocities of each ring using the deviations from the mean velocity and the integrated flux at each point. We carried this out with another two-step procedure, allowing all 77 parameters to be varied to minimize the mean velocity and integrated flux residuals; then we fix the rings' centers and velocities and minimized the overall $\bar{\chi}^2$ as before.

We made one additional modification to the model to better fit the pattern of excess fluxes and deviant velocities that suggested a separate outer ring. The final model allows the outermost 20% of the HI to have a fixed set of parameters that are not interpolated with the values of the parameters at the next radius inward. The minimum $\bar{\chi}^2$ solution is shown by the heavy line in Fig. 5, and the residuals of the are shown in Fig. 6.

Estimating the errors in this case is more difficult because neither the flux nor shape term in equation (3) dominates, so the appropriate range in $\bar{\chi}^2$ is somewhere between 0.058 and 0.011. Also unlike the basic model, changes in the individual parameters affect a much more localized area in the immediate vicinity of the corresponding ring. Therefore we believe it is more appropriate to measure absolute changes in $\bar{\chi}^2$ rather than percentage changes as before. We choose an error interval based on a change in $\bar{\chi}^2$ of 0.04, although this is essentially arbitrary, and even more than before we stress that the errors are only indicative because of correlations between parameters. Also note that the errors in the surface density of HI were actually determined logarithmically.

The best-fit solution has many of the properties we would anticipate based on our discussion in Section 4.2, and it improves the velocity residuals. The outer disk shifts progressively farther to the west-northwest, reaching about 4 arcmin for the outermost rings, and the mean ring velocity becomes $\sim 5 \text{km s}^{-1}$ more negative in the outer regions as the observational data suggest. The model is well-constrained in the bright disk, except in the very center, where the model is essentially unconstrained, as discussed earlier. In the outer regions the fit is somewhat less certain as the variations between models also suggests. This model removes much of the remaining symmetrically patterned residuals and achieves a $\bar{\chi}^2$ of $\sim 12$. It also introduces a distinct velocity component at the right velocity in many of the regions where a bimodal HI profile was observed. However it appears that the gas along this structure is clumpy, so a uniform ring cannot match the fluxes and mean velocities in detail locally. In other extended HI disks, detailed synthesis observations have shown that apparent outer rings may consist of distinct high velocity cloud complexes (for example, Kamphuis & Briggs 1992). The remaining residuals do not show systematic behavior that would be readily modeled by additional rings, but would instead require a more complex model that allows azimuthal variations along the rings, non-circular motions, or clumps of gas following distinct orbits.

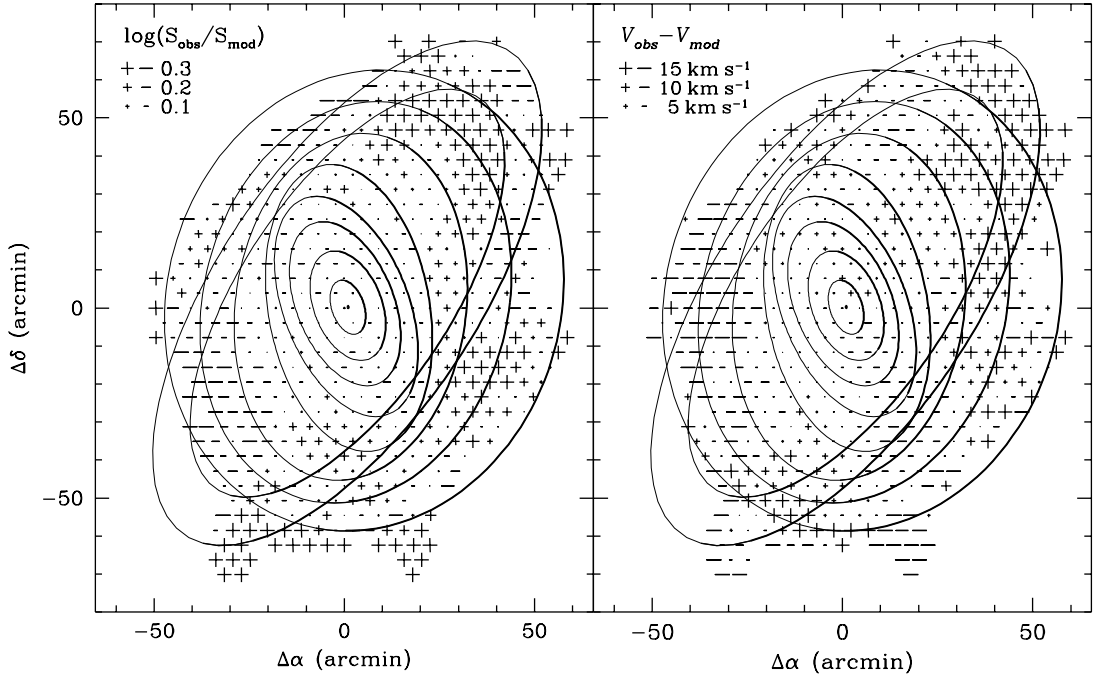

**Figure 6.** Flux residuals for the free ring model described in the text, as in Figure 4.

### 4.4) Tidal perturbations from M31

Some of the asymmetries observed in M33's HI distribution may be related to interactions with its massive neighbor M31. The angular separation between M33 and M31 is about $15.4°$, and they are at similar distances, so they are separated by only $\sim 200$ kpc. The mass of M31 is $\sim 3.5 \times 10^{11}$ $M_\odot$, and from our data we estimate M33's mass is $\sim 10$ times lower: $\sim 3.8 \times 10^{10}$ $M_\odot$, which is about a factor 2 higher than previously estimated.

The ratio between M33's mass and the diameter of its total HI extent cubed, is five times bigger than the ratio between twice the mass of M31 and the distance between M31 and M33 cubed. Therefore M33 is well outside the Roche limit of M31 and only weak tidal effects should disturb the gas distribution in M33. M31 is located northwest of M33 at a position angle of −43°. This is similar to the direction toward which the outer disk of M33 is twisted which could suggest tidal origin.

Weak tidal disturbances are normally symmetric, but the residuals of the intensity in Fig. 4 (as discussed earlier), show an excess of material on the west–northwest side of M33. This could be gas influenced by ram pressure as M33 moves through a possible intergalactic medium within the Local Group, but a more likely cause is non-circular motion.

Observations of lopsided HI profiles in isolated galaxies suggest that such asymmetries might be due to material on elliptical orbits. However, these asymmetries should smear out within a few rotational periods unless external perturbations or peculiar dark matter distributions are present (Baldwin, Lynden-Bell & Sancisi 1980). For M33, it is possible that the gas is orbiting in a non-axisymmetric way due to the distortion of the equipotential surfaces toward M31.

Tidal forces become progressively more asymmetric at larger size scales. Thus for the outermost envelope of M33, at a radius of ∼1/12th of the separation between M33 and M31, the equipotential surface is offset towards M31 by about 4 arcmin, while the shift is < 1 arcmin inside the edge of the M33 bright disk. This is only a rough estimate since the mass of the two objects are not point like and there are some uncertainties in masses and distances as well. However it is worth noticing that the 4 arcmin center shift is similar to what the free ring model suggests for the outermost orbits, in roughly the direction of M31.

## 5. M33 and the Tully-Fisher Relation

Since M33 is one of only a few local calibrators of the extragalactic distance scale, it is important to understand how its warped disk might affect estimations of the galaxy's line width, which is basic to the Tully-Fisher distance method. In an earlier study, Sandage & Humphreys (1980) pointed out that the inner regions of M33 show sharp twisting of the optical isophotes which already indicated potential problems in determining the inclination angle and therefore in deprojecting the rotation velocity from the 21 cm line width. The warp was also modeled by Briggs (1990), but our extended observations trace gas farther out, giving a more complete picture of the kinematics. The strong warping we find has implications for two related questions: (1) How are 21 cm line width measurements of M33 affected by the warped outer disk? and (2) How much could the line width vary in similar galaxies where we had only a single global HI measurement of the entire galaxy? We address these questions in this and the following section.

The first question is in many ways the less interesting, even though M33 is one of the fundamental calibrators of the Tully-Fisher method. Previous observations have already mapped HI in the bright inner regions of the galaxy (Rogstad et al. 1976; Reakes & Newton 1978) and our line widths for the inner disk of M33 are in good agreement with them. It is therefore appropriate to use an inclination correction to the 21 cm line width based on the optical inclination. Furthermore, since the inclination of the outer and inner disk regions happen to be similar, the effects on the 21 cm line width are small—the line width of the entire HI distribution is only ∼3% bigger than that of the inner disk alone.

However, the Tully-Fisher method is more typically applied to galaxies observed at such large distances that the gas is all detected within a single telescope beam. The geometry and inclination of the outer gas is unknown, although it may affect the line width measurements. Since M33 is an archetype for the Tully-Fisher method and we have good information on the behavior of its outer disk, we ask how the line width measurements of other "M33's" might be affected if their outer disks were oriented in an unknown way to the line of sight. A tilted outer disk can cause the overall line width to look broader or narrower than expected from the inner regions alone. It may also cause systematic effects when the inner disk is either face-on or edge-on, since the outer disk would then always have a more moderate inclination. We will examine the scatter and systematic differences that arise as well as the sensitivity of different line width measuring techniques to the tilted outer disk.

We use our basic model of M33 as described by equation (5) but oriented at random in space. We find the line width for the entire galaxy and form its ratio with the line width arising from the inner regions where the gas has about the same inclination as the optical disk. (We define the inner regions for this purpose to be material orbiting inside a 30 arcmin radius of the model, where there is almost no warping present.) This ratio gives the multiplicative error in the observed line width, and it indicates what would be an even larger error in the deprojected rotation speed after corrections are made for the velocity dispersion of the gas.

In Fig. 7 we show the ratio of width measurements determined using several common techniques. In the top panel we show the widths measured at 50% of the peak of the horn on each side of the HI profile as a function of the inclination (Schneider et al. 1986 discuss why this technique should have the best statistical properties). The other three panels of the Figure show the relative widths measured at 20% of peak, and at 50% and 20% of the mean flux in the profile. The "mean" measurements, adopted by a number of authors, are based on the velocity extremes measured at some fraction of the mean flux determined over a window where emission is detected (Bicay & Giovanelli 1986 discuss this line width measurement technique).

The Figure shows that the line widths often become larger than they would be without a warped outer disk. The effect is stronger at smaller inclinations of the inner disk because the outer disk is severely enough tilted that it can remain highly inclined. The effect is also stronger for measurements made at lower fractional levels in the profile because the outer disk produces less 21 cm emission than the inner disk, so the added emission affects the base of the profile more. All of the techniques show prominent effects for more face-on "M33's," and the resulting error in the deprojected rotation speed would dominate over modeling uncertainties in the internal dispersion or measurement uncertainties in the line width or inclination. The widths measured as a fraction of the peak emission appear less affected by the outer disk gas than those measured at a fraction of the mean flux. The "peak" measurements are also somewhat less systematically biased because they can vary up or down depending on whether the outer disk is at a higher or lower inclination than the inner disk: a higher inclination widens the outer part of the profile while having little effect on the peak flux density; a lower inclination adds to the peak flux without contributing to the outer part of the profile, so the measurement is effectively made at a higher (narrower) part of the profile.

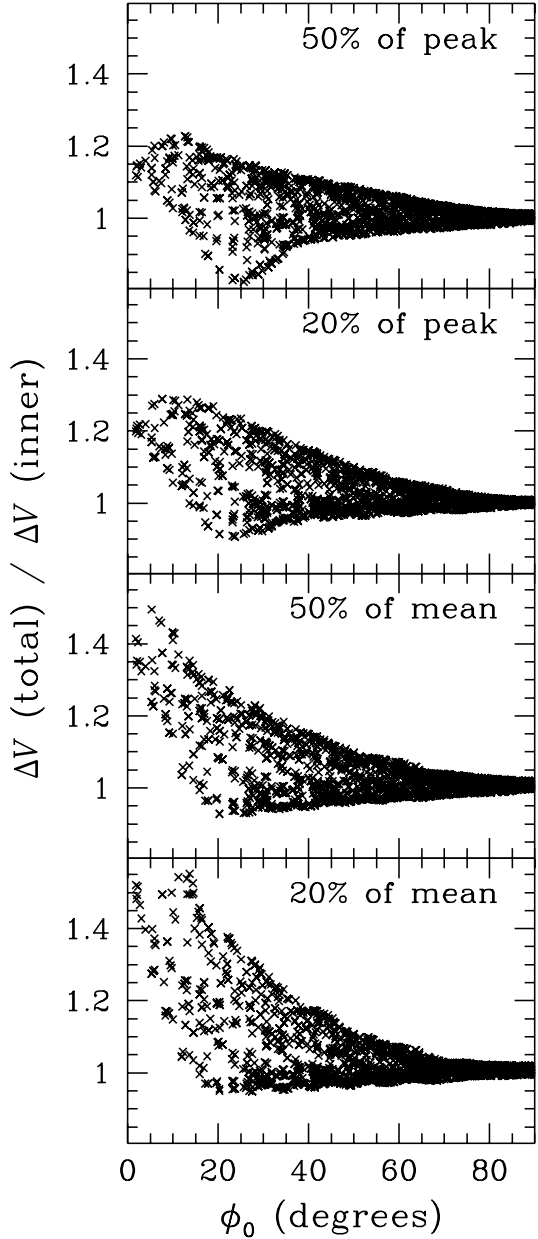

**Figure 7.** The line width ratio between total HI profile and the inner disk alone for galaxies warped like M33 relative. The four panels show results for four measurement techniques described in the text.

The larger deviations seen in the "mean" measurements are mainly a consequence of the lower levels on the profile being measured since the mean is always lower than the peak, which leads to the problem discussed above. Moreover, the lower level in the mean is made even lower when the inner galaxy is near face-on because the mean measurements' velocity window over which the mean is determined is expanded by the presence of the outer gas. The "mean" measurements do not drop below the value for the inner disk when the outer disk is at a lower inclination than the inner disk because the velocity window and integrated flux remain unaffected.

Without any knowledge of the outer disk inclination—as is normally the case for a distant galaxy—the value of $\Delta V$ can be significantly in error at low inclinations. The mean ratio of the line widths and their standard deviations are given for different inclination ranges in Table 1. If we only consider points with inclinations larger than 40° the mean ratio of the line widths are very close to unity, and their standard deviations are quite small. For these higher inclination galaxies, it does not appear to matter much which method is used. However, by using the 50% of peak method, accurate results can be achieved down to lower inclinations, where extinction effects are also less of a problem for optical flux measurements.

If one were to apply typical values for the Tully-Fisher relation (see, for example, Freedman 1990) to a profile width with a 5% error, the predicted magnitude would in turn be in error by ∼0.2 mag. The measurements at low fractional levels of the mean can be much more significantly in error for low inclination galaxies, leading to underestimates by more than one magnitude for individual galaxies and systematic overestimates of the galaxy distances by more than 10%. We recommend adopting the 50% of peak measurement technique to minimize the effects of warped outer disks as well as to improve the statistical accuracy of the measurements.

Up to this point we have assumed that all galaxies have a warp similar to M33, but viewed from random lines of sight. As a complementary case we consider the effect of warps of different amplitude in galaxies with the same inner inclination as M33. In the basic model $\phi_\infty$ does not differ from $\phi_0$ by much and therefore the total HI line width is close to the width of the HI line of the inner part only. Variations of $\phi_\infty$ and $\theta_\infty$ has the following effect almost independently from the way the width is measured: if the warp is such that the outer part is closer to being face-on than the inner part, then the ratio of outer to inner line width is only slightly smaller than one. As $\phi_\infty$ varies from $\phi_0$ to 90 degrees, the ratio increases from 1 to 1.2. If the warp occurs further out than the edge of the bright disk, variations of the total profile width are less drastic, as expected, since the intensity is lower and affects a smaller part of the galaxy.

## 6. Line widths of Other Galaxies

The physical basis of the Tully-Fisher method lies in the relation between the luminous mass and the rotational velocity. The use of inclination-corrected line widths based on the global HI profile introduces an error that is made worse by the presence of warps like that found in M33. In this section, we explore the extent to which published line width data may be consistent with warping as a widespread phenomenon among spiral galaxies, and we discuss possible implications for Tully-Fisher studies. We do not attempt a full-scale study including other sources of error, effects of the propagation of errors, or intrinsic scatter in the Tully-Fisher relation. We refer the reader to Rhee (1996) for a detailed examination of these effects.

Far too few narrow HI profiles are observed among face-on spiral galaxies (Lewis 1987), indicating that spirals fail to obey a simple geometric relationship between the line width and apparent inclination based on the optically observed axis ratio. A variety of models have been proposed to explain this difference, including turbulent broadening, kinematic distortions, warps, bending waves, etc. (see discussion of Lewis 1987). Without a clear understanding of the causes of this line broadening, it is not possible to correct the line widths, and as a result low-inclination galaxies have been excluded from most Tully-Fisher studies. Even if this is considered an acceptable sacrifice, it is important to understand the underlying cause of this broadening since it will also affect line widths at moderate inclinations.

Several attempts have been made to correct low-inclination line widths empirically. For example, Bottinelli et al. (1983) simply subtracted an estimate of the turbulent line width $w_t$ from the observed line width $w_{obs}$ to determine the linewidth $w_c$ due to circular rotation alone:

$$w_c = \frac{1}{\sin \phi_0} \left( w_{obs} - w_t \right) \tag{8}$$

where $w_t$ was modeled on the basis of the fraction of the peak flux density being observed using a geometric model of the inclination dependence, and it was normalized using Tully-Fisher luminosity estimates. However, a linear subtraction is not the correct form for the contribution of turbulent motion to the line width, as Tully & Fouque (1985) point out for dwarf galaxies. In fact, even for large-line-width galaxies, turbulence adds essentially in quadrature for measurements made at 50% of the peak flux density. In other words, it is more accurate to write

$$w_c = \frac{1}{\sin \phi_0} \left( w_{obs}^2 - w_t^2 \right)^{1/2} \tag{9}$$

where all of the widths are measured at 50% of peak. We tested this using realistic line profile shapes with a wide range of line widths and with velocity dispersions ranging from 5 to 25 km s$^{-1}$ (FWHM from 12 to 60 km s$^{-1}$), and equation (9) was accurate to within 5 km s$^{-1}$. At lower points on the HI profile, turbulent motion also requires a linear correction, which depends sensitively on the line width, velocity dispersion, and profile shape. Avoiding this complication is another advantage to using the 50%-of-peak measurements.

Turbulent motion cannot generate the large line width corrections at small inclinations that Bottinelli et al. (1983) find are empirically necessary to match their data. On the other hand, line broadening due to a warped outer disk has a more nearly additive effect on the line width than turbulent broadening, and thus could be an alternative explanation for the effects modeled by these authors. It is also interesting that the corrected line widths of Bottinelli et al. (their Fig. 10) show a divergence at small inclinations much like the deviations caused by warping in our Fig. 7.

Rhee and Broeils (1996) show that a significant reduction in the scatter of the Tully-Fisher relation is achieved when two-dimensional velocity information is used used instead of random-motion corrected line widths. They further find that the random motion correction would have been inappropriate if applied in individual cases, and that the reduction in scatter is not obtained using one-dimensional rotation curves. This underlines again the importance of understanding the underlying kinematics before attempting to interpret HI line widths, which comprise the vast majority of the available data.

If many galaxies have outer disk warping like M33, as is suggested by the study of Briggs (1990), then this could produce an important contribution to their line widths when they are observed at relatively low inclinations. Unfortunately, most previous studies that might have shed some light on the frequency of this phenomenon (for example, Mathewson et al. 1992; Bernstein et al. 1994) were limited to galaxies with high inclinations and excluded those with shallower slopes on the edge of their HI profiles, which is a possible feature of a warped HI disk.

We can test whether HI line widths are affected by warping by studying a large sample of galaxies at all inclinations, and all of similar type and mass so that the rotation speeds and internal velocity dispersions should be similar. We use the extensive HI data available in the literature to provide both line widths, and by converting total HI fluxes to masses we also narrow each type-class according to HI mass to generate a sample with more homogeneous masses.

We have examined references drawn from the bibliographic catalog of Huchtmeier & Richter (1989) for spiral galaxies classified by Nilson (1972) in the UGC. We use only high-quality width measurements made at Arecibo or Green Bank for consistency. Furthermore, we have restricted our comparisons to observations made with sufficient spectral resolution ($\sim 20$ km s$^{-1}$ or better) that instrumental broadening is minor even for narrow profiles.

First we examine the most "M33-like" galaxies with the necessary HI measurements. Our sample consists of 218 Sc through Sd galaxies (M33 is classified as Scd in the UGC) with measured HI masses in the range $9.0 < \log M_{HI} < 9.5$ (M33 has an HI mass of $\log M_{HI} \approx 9.25$). Three quarters of these galaxies have measurements made at 50% of the peak flux density. For the remainder we adjusted the line width according to the mean difference from the 50%-of-peak width: after eliminating outliers, measurements made at 20% of peak were 22 km s$^{-1}$ wider on average, and those made at 50% of the mean were 18 km s$^{-1}$ wider. This linewidth correction is independent of inclination (also see Lewis 1987). We further restricted the comparison galaxies to have extinctions smaller than 0.5 (according to Burstein & Heiles 1982), redshifts in the range $1000 < v_0 < 7000$, and we excluded galaxies within 20° of the Virgo cluster. Galaxies meeting these requirements should be well-classified, and their distances should be fairly well determined from the Hubble law. They are also neither too nearby to fit within the radiotelescope's beam, nor too far away to be detected at a high signal-to-noise ratio. In Fig. 8(a) we plot the observed line widths against the axial ratios ($b/a$) of each galaxy. The Figure also shows the average line width and the error of the mean in equally spaced intervals of the axial ratio.

The surprising aspect of these data is the large number of nearly face-on galaxies ($b/a \approx 1$) that have large line widths. We can quantify the peculiarity of the face-on line widths by comparing $\bar{w}_{face-on}$, the average line width of the nearly face-on galaxies ($b/a \geq 0.9$), with the line width due to circular rotation. Since the 50%-of-peak line widths of relatively edge-on galaxies are only weakly sensitive to the inclination and turbulent velocity, we can estimate $w_c$ by averaging the line widths of all galaxies with $b/a < 0.3$ corrected by $\sin \phi_0$. What we find for the M33-like galaxies is that $\bar{w}_{face-on}/w_c = 0.55 \pm 0.04$.

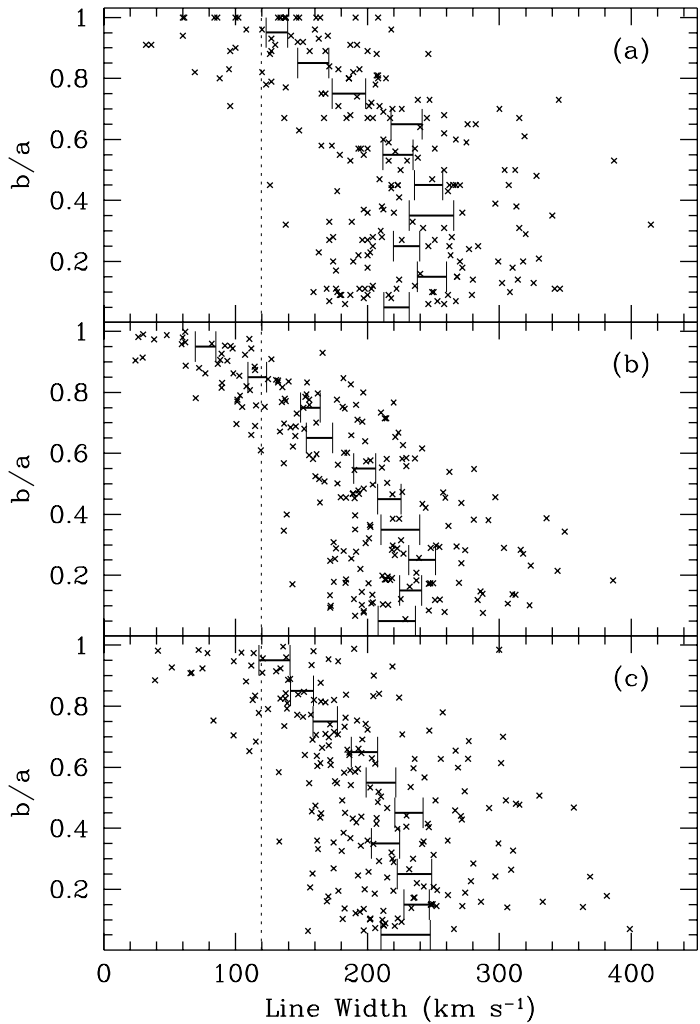

**Figure 8.** (a) The observed line width versus axial ratio for Sc–Sd galaxies like M33. (b) Results of a Monte-Carlo simulation assuming M33-like galaxies with no warp. (c) Results of a Monte-Carlo simulation assuming a distribution of warps with a standard deviation of 25 degrees. The vertical dotted line in all three panels is at $\frac{1}{2}w_c$, and is used to point out the deficiency of narrow-line galaxies at any axial ratio.

A face-on line width as large as $0.55w_c$ cannot be produced by any plausible contribution from turbulent motion or measurement errors as we shall show next. We note first that if the galaxies followed the expected geometric dependence on axial ratio, including the effects of line broadening due to velocity dispersion in the disk and an intrinsic disk thickness of 5%, then $\bar{w}_{face-on}/w_c$ would be 0.31, mostly due to the range of axis ratios included within our "face-on" class. Measurement errors in the axial ratio and the range of masses intrinsic to the sample also affect $\bar{w}_{face-on}/w_c$, but in ways that are difficult to determine analytically. Therefore we carry out Monte-Carlo simulations of the distribution of line widths vs. axis ratios to understand how the observed distribution might have been generated.

In Fig. 8(b) we show how line widths behave relative to the axis ratios using a Monte-Carlo simulation in which we have included realistic parameters for the galaxies and measurement errors. We provide our model galaxies with the same mean edge-on rotation speed as in the real data, along with a range of dynamical masses matching the half-decade range of HI masses. The resulting distribution of line widths at high inclinations is in excellent agreement with the real data. We have also included a random error of 0.06 in the UGC axis ratios as determined by Guthrie (1992), and a $z$-dispersion of 10.4 km s$^{-1}$ (with radial and tangential dispersions 1.5 times larger), which was the best-fit value found for our M33 data. The result is only to increase the line ratio $\bar{w}_{face-on}/w_c$ to 0.35. (Note that the figure shows a realization with the same number of galaxies as in the actual sample, but statistics quoted are based on Monte-Carlo simulations with 10,000 galaxies.)

These large face-on line widths are not confined to just late-type galaxies. Using line widths from the Huchtmeier & Richter catalog (again limited to high-quality measurements made at Arecibo or Green Bank) for galaxies selected with the same HI mass range and other properties as above, the ratio of the width of the face-on disks relative to that of the edge-on disks are as follows:

| type: | S0–Sa | Sab–Sbc | Sc–Sd | Sdm–Sm |
|---|---|---|---|---|
| $w_c$ (km s$^{-1}$): | $354 \pm 40$ | $330 \pm 30$ | $240 \pm 6$ | $190 \pm 8$ |
| $\bar{w}_{face-on}/w_c$: | $0.70 \pm 0.11$ | $0.54 \pm 0.06$ | $0.55 \pm 0.04$ | $0.58 \pm 0.04$ |

where $\bar{w}_{face-on}/w_c \approx 0.35 \pm 0.01$ is the expected ratio for unwarped disks considering the rotation speeds and plausible ranges of disk thickness and velocity dispersion. Overly large values of the face-on line width are seen for other mass ranges within each type as well. The apparently face-on galaxies always had at least $0.4w_c$ in every half-decade mass range of galaxies we examined.

The large line widths seen in face-on galaxies cannot be explained by non-circular orbits within the plane since these do not contribute to the face-on line width. This leaves several possibilities: (1) the line width measurements may be incorrect; (2) the axis ratios may be overestimated; (3) turbulent broadening is larger or less isotropic than we have assumed; or (4) the inclination of some of the HI may be larger than the inner disk implies, that is, warping.

We cannot completely rule out errors in the line widths, however this appears unlikely. The data are restricted to resolutions $< 20$ km s$^{-1}$ which is too fine to account for the broadening. In addition, the 50%-of-peak values we are using are less susceptible to statistical broadening than other values, as discussed in the previous section, and they are less likely to be affected by weak confusion from neighboring sources (Schneider et al. 1986). One might worry that morphological types are difficult to determine for edge-on galaxies (perhaps leading to an underestimate of $w_c$), but our samples show line widths nearly consistent with a simple $\sin\phi_0$ dependence over the entire range $0 < b/a < 0.5$. We also carried out comparisons where we limited the galaxies to those that had more than one line width measurement that were in good agreement. The number of galaxies in these samples generated weaker statistics, but they showed the same large face-on line width peculiarity.

The second possibility would require the axis ratios in the UGC to be in error by more than $\sim 0.2$ (1$\sigma$) which is much larger than comparisons to detailed photometric measurements indicate. However, even supposing the UGC axis ratios were this inaccurate, would not generate the observed line width/axis ratios distribution. Errors in the axis ratio only shift points in Fig. 8 up or down, but there are too few galaxies with narrow line widths at *any* measured axis ratio: For a 10.4 km s$^{-1}$ $z$-dispersion with no warping, $\sim 16\%$ of the galaxies should have line widths smaller than $\frac{1}{2}w_c$; however, only 7% of the linewidths that are observed to be this small. This can be seen from the number of points to the left of the dotted line at $\frac{1}{2}w_c$ in Fig. 8. The low fraction of narrow-line galaxies observed is found despite concentrated efforts by Lewis (1987) and others to make high-quality measurements of narrow-line galaxies.

The possibility that turbulent motions could explain the large line widths of face-on galaxies appears to be ruled out. The 10.4 km s$^{-1}$ dispersion we use in our models is already somewhat larger than is found in other galaxies, and to make the ratio $\bar{w}_{face-on}/w_c$ as large as 0.55 would require far larger dispersions. Because the dispersion increases the line widths of galaxies at all inclinations, an impossibly large velocity dispersion of $\sim$110 km s$^{-1}$ would be required to make the $\bar{w}_{face-on}/w_c$ ratio so large. Radial or tangential streaming or non-circular orbits increase the edge-on line width and would only make the $\bar{w}_{face-on}/w_c$ ratio smaller. Even if the radial and tangential components of the velocity dispersion were artificially set to zero, we would still need a $z$-dispersion larger than $\sim$ 80 km s$^{-1}$. Vertical infall of gas, like high velocity clouds or other intercluster gas, could have such a high dispersion, but the effect on the face-on line width is small due to the small mass involved relative to the disk gas mass. Besides these apparently insurmountable problems, these effects would produce much more steeply sloped edges on HI profiles than are observed, and the distributions of line widths vs. axis ratios generated by increased dispersion do not look like the observed pattern.

The final possibility, that HI line widths are significantly affected by warped disks, appears viable, and it is the only explanation that reproduces the details of the observed line width/axis ratio distribution. We show this by means of another Monte-Carlo simulation. We produce a population of disk galaxies with inner disks randomly oriented to the line of sight, and outer disks at an angle to the inner disk which is normally distributed with standard deviation $\sigma_{warp}$ (randomly distributed in azimuth). Otherwise the model galaxies are treated just as in the model shown in Fig. 8(b).

A Monte-Carlo simulation assuming a distribution of warp angles with $\sigma_{warp} = 30°$ is shown in Fig. 8(c). Thus, most galaxies would have small warps, but a 30° warp would be fairly common. In this model we simply adopt the larger of the line widths from the inner and outer disks as our "observed" value. In reality the value would be an average depending in a complex way on the relative amount of gas in the two regions, *which would require an even larger average warp to achieve as great a face-on line width.* The simulated sample has an average face-on line width that is $0.54w_c$, which matches the real data to within the errors ($0.55 \pm 0.04$). The simulated sample also has a appropriately small fraction of galaxies with line widths $< \frac{1}{2}w_c$: 9%, versus the 7% observed.

This analysis suggests that M33, with its $\sim$30° warp is not unusual among Scd galaxies or even among other types of spirals. And while the effect of warps is mainly on face-on galaxies, it will be competitive with effects on the line width like non-circular rotation even among edge-on galaxies. With careful attention to the shape of HI profiles, for example, looking for "shoulders" or sloped edges, it may be possible to identify the cases where warps are causing significant broadening. And by using line widths measured at higher fractions of the peak flux density, perhaps as high as the 80%-of-peak level suggested by Lewis (1987), it may be possible to minimize the effects of warping on the line width.

A high frequency of warping also raises the possibility of some subtle problems with selection effects. For example, more edge-on galaxies tend to be detected at large distances because of their higher surface brightness, and these will be less affected by warps than a sample at a smaller distance. This might cause systematic biases in Tully-Fisher distances as a function of distance. Another possibility is that galaxies in higher density environments may have a greater likelihood of warps so their line widths may be overestimated relative to field spirals. In each case the bias may be small, but it could be important for peculiar velocity studies.

## 7. Summary

We have carried out detailed observations of the HI in M33 using the Arecibo radiotelescope. The gas exhibits widely different properties in the inner and outer portions of the galaxy, which we have attempted to model in terms of gas on circular orbits with changing inclinations and rotation speeds. We find that:

(1) M33 has an outer disk containing about one quarter of the total HI oriented at $\sim 30°$ with respect to the inner disk.

(2) The rotation speed appears to be flat or rising slightly out to the limits of our observations at about twice the optical radius usually quoted.

(3) The outer HI distribution is skewed in the general direction of M31, which may be distorting the HI distribution into a slightly asymmetric distribution due to nonlinear tidal effects.

(4) The data suggest the presence of clumps and possibly a weak distinct outer ring of HI, oriented at a large angle to the inner disk of the galaxy.

In addition, we have examined implications of the gas kinematics in M33 for the Tully-Fisher relation, both in the particular case of M33 and more generally for spiral galaxies with warped outer disks of HI:

(5) The line width of M33 varies only slightly depending on whether the entire galaxy or just the inner disk is measured, since the inner and outer disks of the galaxy are fortunately almost at the same inclination to our line of sight. The best measurement for the inner disk alone is 187 km s$^{-1}$ at 50% of the peak flux density, which would be the best measurement to compare to other line widths measured in galaxies without warps.

(6) For galaxies with warped outer disks (which are indistinguishable from unwarped galaxies in most single-dish 21 cm observations), line width measurements at a low fraction of the peak intensity are biased toward artificially large values and have a larger scatter. Nevertheless, by using 50%-of-peak flux density in each "horn" in line width measurements, there should be relatively small effects by warps on the line width down to inclinations of $\approx 20°$.

(7) An examination of available HI data shows that the distribution of line widths with axis ratio is consistent with a model in which outer warps of fairly large amplitude are a common phenomenon in all types and mass ranges of spiral galaxies.

We would like to thank Ed Salpeter for having suggested this project, and the staff of the Arecibo observatory for their assistance in acquiring the data. We also thank an anonymous referee who suggested many valuable improvements to the paper. This work was supported in part by NSF Presidential Young Investigator award AST-9158096.